**Emerging basic, clinical and translational research fronts in dental biomaterials R&D**

David Fajardo-Ortiz,[1] Pablo Jaramillo,[1,2] Claudia Jaramillo,[2] Raul Reséndiz,[2] Miguel Lara-Flores,[2] Victor M. Castano[2]

1. Centro de Investigación en Políticas, Población y Salud, Facultad de Medicina, Universidad Nacional Autónoma de México. Cd. Universitaria, Cd. de México, 04510 MEXICO.

2. Coordinación de Vinculación Institucional, Secretaría de Desarrollo Institucional, Universidad Nacional Autónoma de México Cd. Universitaria, Cd. de México, 04510 MEXICO.

## Abstract

The current (2007-2007) structure and content of dental materials research has been investigated by identifying and describing the emergent research fronts which can be related to basic, translational and clinical observation research. By a combination of network analysis and text mining of the literature on dental materials indexed in the Web of Science, we have identified eleven emerging research fronts. These fronts are related to different dental materials applications which are at different levels in the knowledge translation and biomedical innovation process. We identified fronts related to dominant designs like titanium implants, competing technologies like ceramics and composites applications to prothesis and restauration, and disruptive technologies like nanomaterials and mineral trioxide aggregates. Our results suggest the possible relation between the technological complexity of the dental materials and the level of advance in terms of knowledge translation. This is the first time the structure and content of research on dental materials research is analyzed.

## Introduction

Research in dental materials is a consolidated and complex interdisciplinary field that can be regarded as a trading zone between materials science (i.e. a branch of physics, chemistry and engineering) and dentistry (i.e. branch of medicine).[1-2] When two or more disciplines interact with each other, there is an interchange of concepts and research instruments but, more importantly, there is an intense negotiation process, for these two disciplines exhibit different objectives, approaches and standards that define the membership criteria of their corresponding disciplinary communities.[3-5] In the case of dental materials, this negotiation may be expressed in terms of improving the quality of life of the patient, by developing and implementing biocompatible, long-lasting and aesthetically-satisfactory dental materials.[1] Achieving the former goal requires an effective communication process between basic research (centered on the physical, chemical and biological properties of the materials, as well as the engineering involved) and the clinical practice (i.e., patient-centered).[1, 6-7] During the two last decades, it has been pointed out the need of translating the basic knowledge produced in the laboratory into clinical practice and improvements in the health status of the population.[8-9] In that sense, the emergence of translational transdisciplinary research communities has been proposed as the necessary bridge that connects basic research with clinical practices.[4, 10-11] These transdisciplinary communities would be characterized by the combined use of terms and research instruments from the disciplines that are connected. [4, 10-11] Also, the literature produced by these communities tends to be located in a central position in the citations networks. [4, 10-11] Accordingly, we have developed a successful methodology to study the knowledge translation process in biomedical technologies and the research on diseases. [10, 12-15] This methodology is based on a combination of network analysis and text mining as to identify and analyze the emergence of clinical basic and translational research fronts. [10, 12-15] Research fronts are clusters in the citation networks [16] which are produced by a virtual community of scientist or invisible colleges that work on closely-related problems with similar approaches. Because the modularity (the tendency to form modules, i.e., "regions in the networks within which connections are dense but between which they are sparser") [17] and homophily (nodes with similar attributes tend to attach to each other more frequently than dissimilar ones) [18] are recurrent qualities of complex networks (like social and literature networks), [18, 19] it is possible to identify the research fronts. Additionally, the analysis of the content in regions of the literature networks allows to determine whether those regions are mostly related to basic, translational or clinical research.

Thus, the objective of this present study was to determine the current (2007-2007) organization of the research on dental materials in terms of content and interaction of the research fronts. Particularly, we wanted to identify the emergence of fronts related to basic, translational and clinical research.

## Methodology

The methodology has previously been successfully used to analyze the organization of the research fronts in the study of diseases[12-13] and biomedical nanotechnologies[14-15]:

1. A search of papers on dental materials was performed in the Web of Science[20] during May, 2017. The search criteria were the following: TOPIC: (material or biomaterials) AND TOPIC: (dentistry or odonthology or dental) Timespan: 2007-2017. Indexes: SCI-EXPANDED, SSCI, A&HCI, CPCI-S, CPCI-SSH, BKCI-S, BKCI-SSH, ESCI. 24,046 papers were found.

2. A network model of 17,241 papers and 73,954 inter-citations was built with the papers found in the Web of Science by using the software HistCite.[21] Then, the network model was analyzed and visualized with Cytoscape. [22] A core sub-network of papers with an indegree (inter-citations) ≥ 6 was then closely examined. This sub-network consists of 3,456 papers and 14, 251 inter-citations. Importantly, this sub-set of 3,456 papers received 75% (56,023 out of 73,954) of the citations from the whole network.
3. A Cluster analysis based in the Newman modularity[17] was performed on the core subnetwork using Clust&see, a Cystoscape plug-in. [23] This analysis divided the sub-network of citation in several research fronts (clusters or modules of papers). It is important to mention that the three first step of the methods imply a definition of research fronts that meets the following criteria: research fronts are subsets of papers on a particular topic that are (a) recently published (from 2007 to the present), (b) highly cited (indegree ≥ 6, i.e., the top 20% most cited), and (c) they form sub-networks within which connections are dense but between which they are sparse.
4. The content of the identified research fronts -the abstract of their papers- was analyzed with KH Coder, software for quantitative content analysis (text mining) [24]. A correspondence analysis plot and a list of top ten most distinctive words was generated with KH coder. This analysis provides useful information about the content and relation among the fronts.
5. The papers with the five highest indegree within each of the research fronts were identified and closely revised.

**Results**

A network model of 3,456 highly cited papers on dental materials was built (Figure 1). The cluster analysis of the network identified 13 clusters. However, two clusters were too small to be considered as relevant research fronts. The content and interconnection of the research fronts revealed the organization of the last ten years of research on dental materials (Figures 1-3). According to our results, the research fronts are organized around different types and applications of dental materials (Figures 1-3 and Table 1). However, the most important feature of the organization of the recent research on dental materials is that the research fronts are arranged according to a research level continuum ranging from biomedical basic research to clinical practice. That is, the clustering analysis of the network model identified basic, translational and clinical research fronts. At one end of the continuum there is the most clinical research front (Front 4; Figures 1 and 2), which is related to the evaluation of techniques aimed to provide the osseous tissue for supporting implants (Figures 1-3 and Table 1). The research front 1 follows front 4 into the continuum (Figures 1 and 2). This front is related to the study of the clinical implication of the use, design and placement techniques of implants (Table 1). These two fronts (1 and 4) are closely related in terms of connectivity and content and they are characterized by reporting technique-sensitive outcomes, a patient-centered approach, and the lesser importance of the type of materials used (Figure 3 and table 1). Following these two clinical research fronts, there is a set of translational research fronts (Fronts 2, 5 8, 10 and 11) that are characterized by the study of the clinical impact, and more importantly, the performance of dental materials (Figures 1-3 and Table 1). Front 2 is related to the study of the performance and clinical impact of different materials for dental prostheses. Front 5 is related to the study of clinical performance of restorative materials mainly resin-based, amalgam and composites (Table 1). Front 8 is aimed to the evaluation of alternatives to titanium implants, particularly zirconia implants (Table 1). Front 10 is a transition zone between front 2 and 5 (Figure 1) and its content is too diverse as to be considered a proper research front (Table 1). Front 11 is related to the evaluation of methods of manufacturing dental prostheses with higher accuracy fit (Table 1). Because its content and connectivity, front 11 could be considered an “appendix” of Front 2 (Figures 1-3 and table 1). Finally, at the opposite end of the continuum there is a set of basic research fronts (Figure 1 and 2) focused on the description (Front 3 and 6) and explanation (Front 7 and 9) of the properties of dental materials (Figure 3 and Table 1). Front 3 is basic research on the mechanical properties of dental materials, particularly composites (Table 1). Front 6 is research aimed to the study of antibacterial properties of dental materials (Table 1). Front 7 is Basic research on the physical-chemical and biological properties of mineral trioxide aggregate and the mechanism that explain these properties (Table 1). Front 9 is related to the study of the cytotoxicity of resin monomers, mainly tetraethyleneglycol dimethacrylate (TEGDMA) and 2-hydroxyethyl methacrylate (HEMA), particularly the investigations on the biomolecular mechanism of cytotoxicity (Table 1).

One interesting result is the particularly notorious homophily in the network model, as it can be observed when we compared the structure of the network model with the clustering of the fronts in the correspondence analysis. That is, closer research fronts tend to have similar contents.

## Discussion

In order to understand the relevance of these results, it is important to keep in mind that, while dental materials research is a highly consolidate discipline -with more than a century of history, if we consider the “amalgam war” from 1840 to 1850 as a starting point-[1] the scope of this study is limited to the last decade (2007-2017). In the present investigation, we wanted to know how the recent research on dental materials is structured. By focusing on the recent papers, we obtained a view on the level of the technological development of different types of dental materials and their application. That is, our study identified research fronts related to consolidated technologies (the de facto standards or dominant designs[25]) like titanium implants; competing technologies (in which there is no de facto standard[25] like restorative materials and materials aimed to make prostheses, and potentially disruptive technologies like nanomaterials with antibacterial properties. [26] Interestingly, our results suggest that, when a dental technology becomes the dominant design (like titanium implants in research fronts 1 and 4), the relevance of its properties is less important than its implementation in the clinical practice. In that sense, the key variables in these investigations (clinical observations) are the clinical procedures and the skills of the dental practitioners. In the case of competing dental materials (front 2, 5 and 8 that are related to translational research), neither properties of these materials nor skills and procedures are at the core of the discussion. Instead, the clinical performances of the different materials are the main topic in these fronts. Finally, in the case of potentially disruptive dental materials the description and explanation of the materials is the central topic (basic research). In a previous investigation, we have studied the knowledge translation process in liposomes as a technology used to treat cancer by analyzing the structure and content of the literature networks. [15] In that study, we found a similar organization of the investigation on liposomes and cancer: basic research fronts associated to the development of disruptive technologies (the use liposomes in hyperthermia therapy, gene therapy and the combination of liposomes with small interfering RNA) and translational and clinical research front related to the development and implementation of the drug Doxil (liposomal Doxorubicin) which is the dominant design in cancer nanotechnology. [15] The most relevant observation of that study was that among all the possible alternatives using liposomes the most successful in terms of the knowledge translation process was the combination of liposomes with old small molecules drugs. [15] That is, there is a trade-off between the complexity of a technological alternative (which increase the time, effort and expense of development) and the increase in performance [See 27]. This trade-off would, in combination with historical circumstances (which type of dental components and materials where used first) offer a partial hypothesis on the structure and content of the emerging research fronts in dental materials research. That is, the level of complexity of titanium as a material for implants could be enough low to allow researchers progress further and focus on the study of its implementation, whereas ceramics and composites represent a higher level of complexity. Dental nanomaterials would represent another increase in the complexity and this could delay its evolution in the knowledge translation process. On the other hand, mineral trioxide aggregate (MTA) has been pointed out as a disruptive and successful innovation. [28] In that sense, it is important to notice that front 7, which focused on MTA research, stands out of the crowd in terms of content (correspondence analysis) and network connectivity as it can be observed in figures 1 to 3. MTA research is strongly basic because of the need to explain the mechanism of interaction but also because its boosting of imitation research and development. That is, the development of variants of MTA and materials with similar properties. Finally, basic research in front 9 is also different from the rest of the fronts because it is not related to the development, testing or implementation of dental materials but it is focused on the explanation of the toxicity produced by some resin monomers. Further research is required to determine the extension of and explain the relations between the complexity of dental materials and their evolution in the knowledge translation process. Also, it would be particularly interesting to study MTA as an example of successfully disruptive technology.

To the best of our knowledge, this is the first time the structure and content of dental materials research has been analyzed. We found research fronts arranged along of a translational continuum, ranging from basic to clinical observation research. The fronts are related to different dental material technologies at different stages of the innovation process. There are clinical research fronts related to titanium dental implants which could be considered an example of dominant design. In a middle position in the network, and in the correspondence analysis plot, we identified translational research fronts in which different types of materials compete with each other. Finally, we found basic research related to the description and explanation of interaction mechanism between tissues and experimental new materials. Among the research fronts stands out the front related to MTA research, which could be considered an example of disruptive innovation. We speculate on the possible relation between the technological complexity of the dental materials and their evolution in the innovation and knowledge processes. Further investigation on this matter would be fundamental to understand the development trajectories of biomedical and dental technologies.

**Author contributions statement**

DF-O designed the research, performed the network analyses and text mining and wrote of the paper. PJ, CJ and RR built the database. All authors contributed to the interpretation of results. All authors read and approved the final manuscript.

**Figure 1.** The network model of highly cited papers on dental materials. The model is displayed by using the "yFiles organic" algorithm. The color of nodes (representing the papers) indicates which research front they belong to. The vertices represent the citations. Note that the research fronts are ranked according to the number of papers that constitute them.

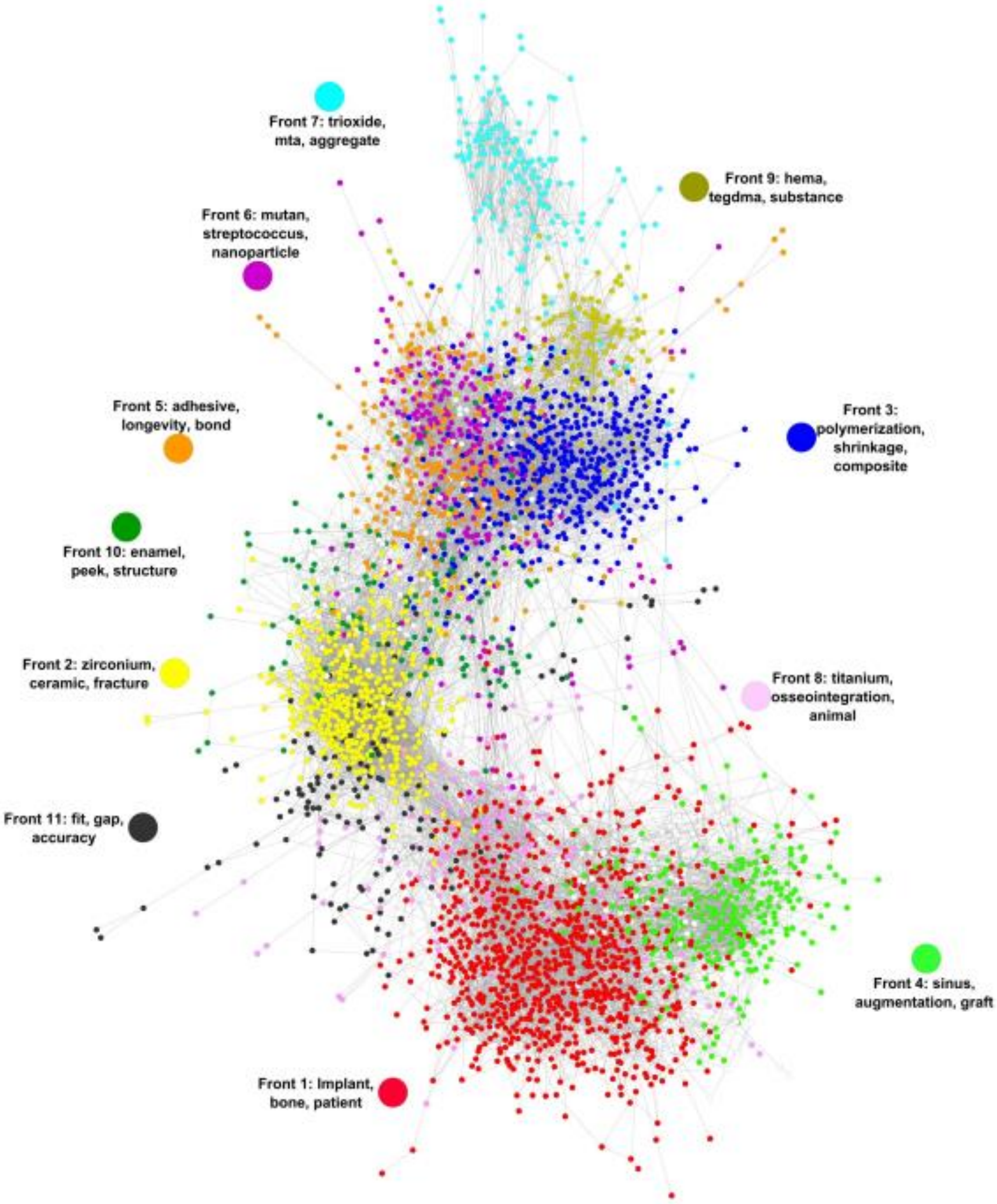

**Figure 2.** Main interactions among the research fronts. Each node represents one of the seven research fronts. The edges represent the sum of the inter-citations between two clusters. Only the main interactions among the fronts are shown.

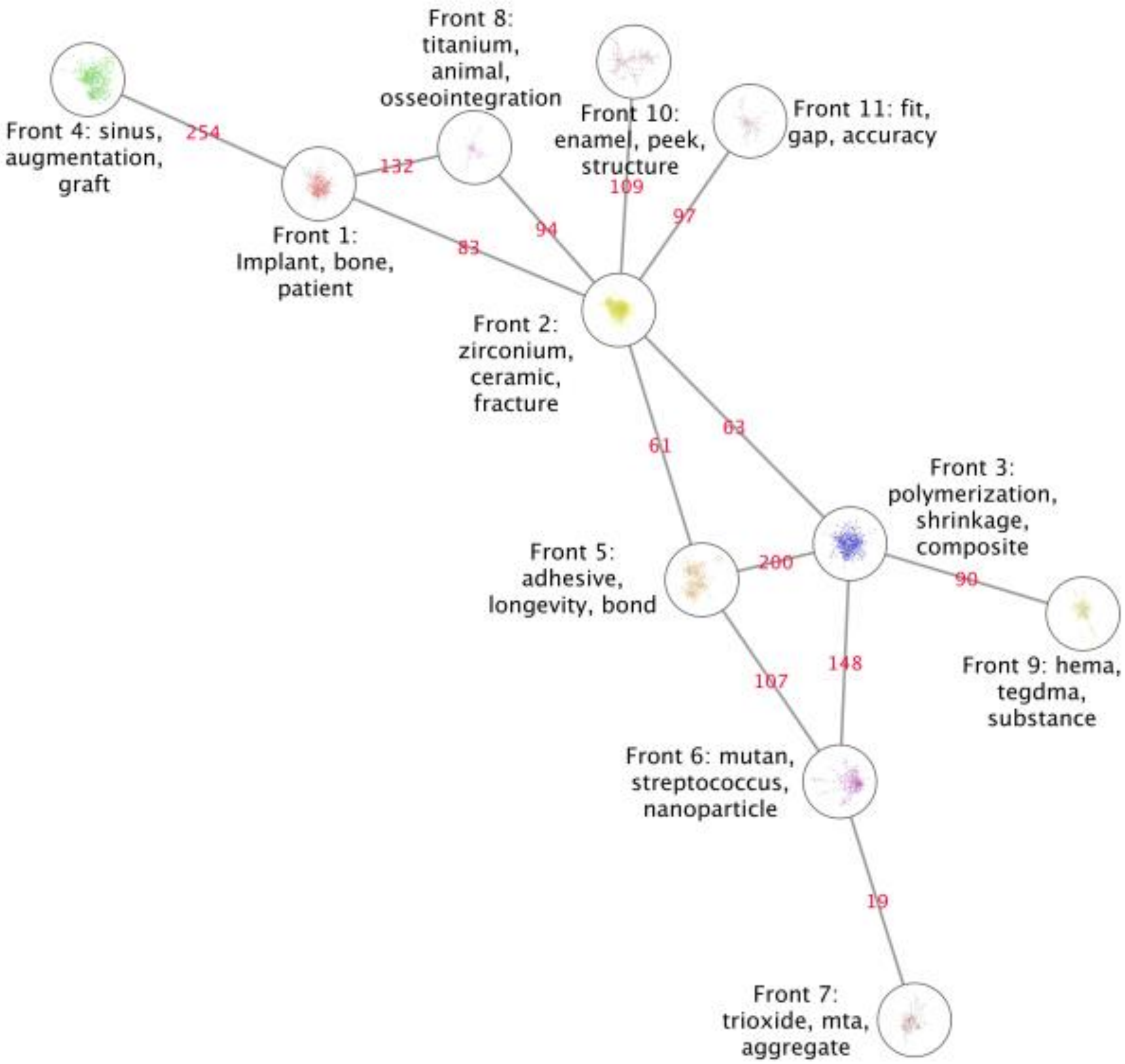

**Figure 3.** KH Coder correspondence analysis of the content of the research fronts. The distance among the research fronts in the plot is a measure of the similarity of their content. The top 50 words filtered by their chi-square value were plotted. The size of the squared bubbles is related to the number of papers of each research front. The size of the round bubbles is related to the number of times a term appear in the papers.

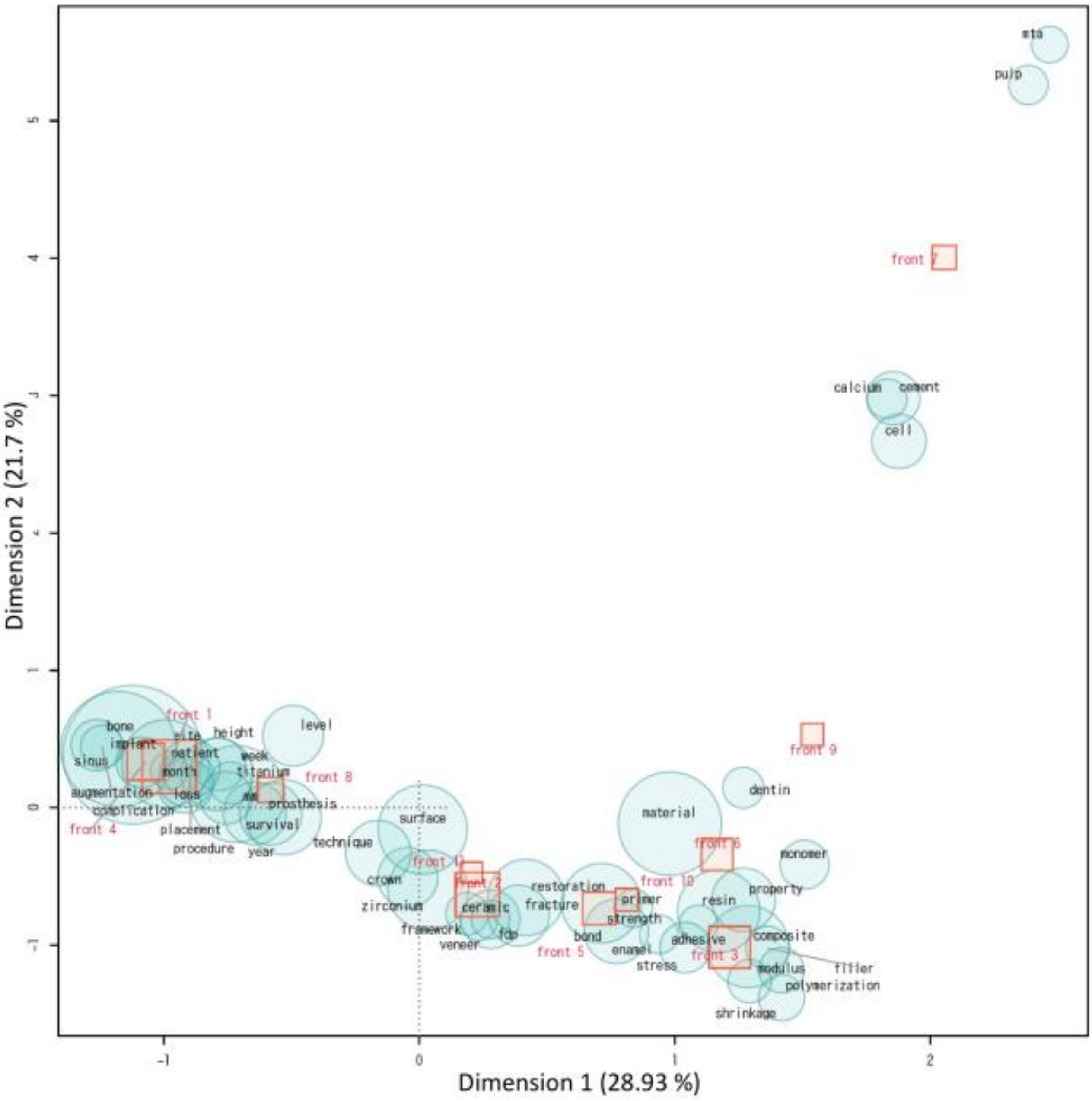

**Table 1.** Description of the research fronts: number of papers, most distinctive terms (Jaccard index) and papers with the highest indegree.

| FRONT 1 | Size: 879 papers and 2,982 inter-citations |
|---|---|
| Top papers: | |
| Zitzmann NU, 2008, J CLIN PERIODONTOL, V35, P286 | Definition and prevalence of peri-implant diseases |
| Canullo L, 2007, INT J ORAL MAX IMPL, V22, P995 | Preservation of Peri-implant Soft and Hard Tissues Using Platform Switching of Implants Placed in Immediate Extraction Sockets: A Proof-of-Concept Study with 12- to 36-month Follow-up. |
| Berglundh T, 2007, CLIN ORAL IMPLAN RES, V18, P147 | Bone healing at implants with a fluoride-modified surface: an experimental study in dogs |
| KH Coder terms: implant, bone, placement, loss, mm, patient, surgery, month, change | |

| FRONT 2 | Size: 463 papers and 2,474 inrer-citations |
|---|---|
| Top papers: | |
| Denry I, 2008, DENT MATER, V24, P299 | State of the art of zirconia for dental applications |
| Sailer I, 2007, CLIN ORAL IMPLAN RES, V18, P86 | A systematic review of the survival and complication rates of all-ceramic and metal-ceramic reconstructions after an observation period of at least 3 years. Part II: fixed dental prostheses |
| Pjetursson BE, 2007, CLIN ORAL IMPLAN RES, V18, P73 | A systematic review of the survival and complication rates of all-ceramic and metal-ceramic reconstructions after an observation period of at least 3 years. Part I: single crowns |
| Sailer I, 2009, INT J PROSTHODONT, V22, P553 | Randomized Controlled Clinical Trial of Zirconia-Ceramic and Metal-Ceramic Posterior Fixed Dental Prostheses: A 3-year Follow-up |
| KH Coder terms: zirconium, ceramic, fracture, veneer, framework, failure, strength, core, crown, test | |

| FRONT 3 | Size: 451 papers and 1,791 inter-citations |
|---|---|
| Top papers: | |
| Beun S, 2007, DENT MATER, V23, P51 | Characterization of nanofilled compared to universal and microfilled composites |
| Ilie N, 2009, DENT MATER, V25, P810 | Macro-, micro-and nano-mechanical investigations on silorane and methacrylate-based composites |
| Ilie N, 2009, CLIN ORAL INVEST, V13, P427 | Investigations on mechanical behaviour of dental composites |
| Watts DC, 2008, DENT MATER, V24, P1 | Axial shrinkage-stress depends upon both C-factor and composite mass |
| Ferracane JL, 2011, DENT MATER, V27, P29 | Resin composite-State of the art |
| KH Coder terms: Polymerization, shrinkage, composite, filler, modulus, property, stress, resin, degree, filtek | |

| FRONT 4 | Size: 257 papers and 1,035 inter-citations |
|---|---|
| Top papers: | |
| Pjetursson BE, 2008, J CLIN PERIODONTOL, V35, P216 | A systematic review of the success of sinus floor elevation and survival of implants inserted in combination with sinus floor elevation - Part I: Lateral approach |
| Aghaloo TL, 2007, INT J ORAL MAX IMPL, V22, P49 | Which hard tissue augmentation techniques are the most successful in furnishing bony support for implant placement? |
| Tan WC, 2008, J CLIN PERIODONTOL, V35, P241 | A systematic review of the success of sinus floor elevation and survival of implants inserted in combination with sinus floor elevation - Part II: Transalveolar technique |
| KH Coder terms: Sinus, augmentation, graft, complication, floor, height, morbidity, posterior, bone, patient | |

| FRONT 5 | Size: 260 papers and 737 intercitations |
|---|---|
| Top papers: | |
| Breschi L, 2008, DENT MATER, V24, P90 | Dental adhesion review: Aging and stability of the bonded interface |
| Opdam NJM, 2007, DENT MATER, V23, P2 | A retrospective clinical study on longevity of posterior composite and amalgam restorations |
| Bernardo M, 2007, J AM DENT ASSOC, V138, P775 | Survival and reasons for failure of amalgam versus composite posterior restorations placed in a randomized clinical trial |
| KH Coder terms: Adhesive, longevity, bond, class, bonding, dentin, number, performance, caries,amalgam | |

| FRONT 6 | Size: 249 papers and 909 inter-citations |
|---|---|
| Top papers: | |
| Beyth N, 2007, J DENT, V35, P201 | An in vitro quantitative antibacterial analysis of amalgam and composite resins |
| Wiegand A, 2007, DENT MATER, V23, P343 | Review on fluoride-releasing restorative materials - Fluoride release and uptake characteristics, antibacterial activity and influence on caries formation |
| Drummond JL, 2008, J DENT RES, V87, P710 | Degradation, fatigue, and failure of resin dental composite materials |
| Xie D, 2011, DENT MATER, V27, P487 | Preparation and evaluation of a novel glass-ionomer cement with antibacterial functions |
| KH Coder terms: Mutan, streptococcus, activity, nanoparticle, caries, bacterium, growth, ion, silver, property | |

| FRONT 7 | Size: 169 papers and 519 inter-citations |
|---|---|
| Top papers: | |
| Camilleri J, 2007, INT ENDOD J, V40, P462 | Hydration mechanisms of mineral trioxide aggregate |
| Torabinejad M, 2010, J ENDODONT, V36, P190 | Mineral Trioxide Aggregate: A Comprehensive Literature Review-Part II: Leakage and Biocompatibility Investigations |
| Min KS, 2008, J ENDODONT, V34, P666 | Effect of mineral trioxide aggregate on dentin bridge formation and expression of dentin sialoprotein and heme oxygenase-1 in human dental pulp |
| KH Coder terms: Trioxide, Mta, Aggregate, Mineral, Pulp, Hydroxide, Calcium, Cement, Cell, Portland | |

| FRONT 8 | Size: 144 papers and 410 inter-citations |
|---|---|
| Top papers: | |
| Gahlert M, 2007, CLIN ORAL IMPLAN RES, V18, P662 | Biomechanical and histomorphometric comparison between zirconia implants with varying surface textures and a titanium implant in the maxilla of miniature pigs |
| Oliva J, 2007, INT J ORAL MAX IMPL, V22, P430 | One-year follow-up of first consecutive 100 zirconia dental implants in humans: a comparison of 2 different rough surfaces. |
| Andreiotelli M, 2009, CLIN ORAL IMPLAN RES, V20, P32 | Are ceramic implants a viable alternative to titanium implants? A systematic literature review |
| KH Coder terms: Titanium, Osseointegration, Animal, Pig, Ti, Bic, Zirconium, Adult, Removal, slum | |

| FRONT 9 | Size: 139 papers and 425 inter-citations |
|---|---|
| Top papers: | |
| Schweikl H, 2007, DENT MATER, V23, P688 | Inhibition of TEGDMA and HEMA-induced genotoxicity and cell cycle arrest by N-acetylcysteine |
| Samuelsen JT, 2007, DENT MATER, V23, P34 | Apoptosis induced by the monomers HEMA and TEGDMA involves formation of ROS and differential activation of the MAP-kinases p38, JNK and ERK |
| Moharamzadeh K, 2007, J MATER SCI-MATER M, V18, P133 | HPLC analysis of components released from dental composites with different resin compositions using different extraction media |

| |
|---|
| KH Coder terms: Hema, Tegdma, Substance, Compound, Ro, Monomer, Triethylene, Cytotoxicity, Methacrylate, toxicity |

| FRONT 10 | Size: 137 papers and 330 inter-citations |
|---|---|
| Top papers: | |
| He LH, 2008, J MECH BEHAV BIOMED, V1, P18 | Understanding the mechanical behaviour of human enamel from its structural and compositional characteristics |
| He LH, 2007, J DENT, V35, P431 | Enamel—a “metallic-like” deformable biocomposite |
| Stawarczyk B, 2012, CLIN ORAL INVEST, V16, P1669 | Load-bearing capacity of CAD/CAM milled polymeric three-unit fixed dental prostheses: effect of aging regimens |
| KH Coder terms: Enamel, Peek, Structure, Hardness, Indentation, Rocatec, Deformation, Resistance, M, science | |

| FRONT 11 | Size: 131 papers and 272 inter-citations |
|---|---|
| Top papers: | |
| Beuer F, 2009, DENT MATER, V25, P94 | Marginal and internal fits of fixed dental prostheses zirconia retainers |
| Reich S, 2008, EUR J ORAL SCI, V116, P579 | Clinical fit of four-unit zirconia posterior fixed dental prostheses |
| Wettstein F, 2008, EUR J ORAL SCI, V116, P272 | Clinical study of the internal gaps of zirconia and metal frameworks for fixed partial dentures |
| KH Coder terms: Fit, Gap, Accuracy, Casting, Replica, Cast, Silicone, Manufacturing, Magnification, mum | |